# ZipNet-GAN: Inferring Fine-grained Mobile Traffic Patterns via a Generative Adversarial Neural Network

Chaoyun Zhang
School of Informatics
University of Edinburgh, UK
chaoyun.zhang@ed.ac.uk

Xi Ouyang
School of Electronic Information and Communications,
Huazhong University of Science and Technology, China
ouyang@hust.edu.cn

Paul Patras
School of Informatics
University of Edinburgh, UK
paul.patras@ed.ac.uk

## ABSTRACT

Large-scale mobile traffic analytics is becoming essential to digital infrastructure provisioning, public transportation, events planning, and other domains. Monitoring city-wide mobile traffic is however a complex and costly process that relies on dedicated probes. Some of these probes have limited precision or coverage, others gather tens of gigabytes of logs daily, which independently offer limited insights. Extracting fine-grained patterns involves expensive spatial aggregation of measurements, storage, and post-processing. In this paper, we propose a mobile traffic super-resolution technique that overcomes these problems by inferring narrowly localised traffic consumption from coarse measurements. We draw inspiration from image processing and design a deep-learning architecture tailored to mobile networking, which combines Zipper Network (ZipNet) and Generative Adversarial neural Network (GAN) models. This enables to uniquely capture spatio-temporal relations between traffic volume snapshots routinely monitored over broad coverage areas ('low-resolution') and the corresponding consumption at 0.05 km$^2$ level ('high-resolution') usually obtained after intensive computation. Experiments we conduct with a real-world data set demonstrate that the proposed ZipNet(-GAN) infers traffic consumption with remarkable accuracy and up to 100× higher granularity as compared to standard probing, while outperforming existing data interpolation techniques. To our knowledge, this is the first time super-resolution concepts are applied to large-scale mobile traffic analysis and our solution is the first to infer fine-grained urban traffic patterns from coarse aggregates.

## CCS CONCEPTS

•**Networks** → *Network measurement;*

## KEYWORDS

mobile traffic; super-resolution; deep learning; generative adversarial networks

## 1 INTRODUCTION

The dramatic uptake of mobile devices and online services over recent years has triggered a surge in mobile data traffic. Industry forecasts monthly global traffic consumption will surpass 49 exabytes ($10^{18}$) by 2021, which is a seven-fold increase of the current utilisation [1]. In this context, in-depth understanding of mobile traffic patterns, especially in large-scale urban cellular deployments, becomes critical. In particular, such knowledge is essential for dynamic resource provisioning to meet end-user application requirements, Internet services that rely on user location information, and the optimisation of transportation systems [2–4].

Mobile traffic measurement collection and monitoring currently rely on dedicated probes deployed at different locations within the network infrastructure [5]. Unfortunately, this equipment *(i)* either acquires only coarse information about users' position, the start and end of data sessions, and volume consumed (which is required for billing), while roughly approximating location throughout sessions – Packet Data Network Gateway (PGW) probes; *(ii)* or has small geographical coverage, e.g. Radio Network Controller (RNC) or eNodeB probes that need to be densely deployed, require to store tens of gigabytes of data per day at each RNC [6], but cannot independently quantify data consumption. In addition, context information recorded is often stale, which renders timely inferences difficult. Mobile traffic prediction within narrowly localised regions is vital for precision traffic-engineering and can further benefit the provisioning of emergency services. Yet this remains costly, as it involves non-negligible overheads associated with the transfer of reports [5], substantial storage capabilities, and intensive off-line post-processing. In particular, combining different information from a large number of such specialised equipment is required, e.g. user position obtained via triangulation [7], data session timings, traffic consumed per location, etc. Overall this is a challenging endeavour that cannot be surmounted in real-time by directly employing measurements from independent RNC probes.

In order to simplify the analysis process, mobile operators make simple assumptions about the distribution of data traffic consumption across cells. For instance, it is frequently assumed users and traffic are uniformly distributed, irrespective of the geographical layout of coverage areas [8]. Unfortunately, such approximations are usually highly inaccurate, as traffic volumes exhibit considerable disparities between proximate locations [3]. Operating on such simplified premises can lead to deficient network resources allocations and implicitly to modest end-user quality of experience. Alternative coarse-grained measurement crowdsourcing [9] remains impractical for traffic inference purposes.

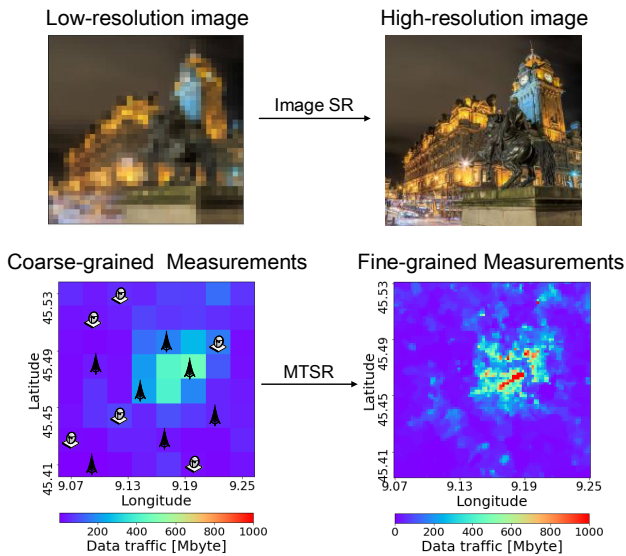

**Figure 1: Illustration of the image super-resolution (SR) problem (above) and the underlying principle of the proposed mobile traffic super-resolution (MTSR) technique (below). Figure best viewed in color.**

In this paper we propose an original mobile traffic 'super-resolution' (MTSR) technique that drastically reduces the complexity of measurements collection and analysis, characteristic to purely RAN based tools, while gaining accurate insights into *fine-grained mobile traffic patterns* at city scale. Our objective is to precisely infer narrowly localised traffic consumption from aggregate coarse data recorded by a limited number of probes (thus reducing deployment costs) that have arbitrary granularity. Achieving this goal is challenging, as small numbers of probes summarise traffic consumption over wide areas and thus only provide 'low-resolution' snapshots that cannot capture distributions correlated with human mobility. Further, measurement instrumentation is non-uniformly deployed, as coverage area sizes depend on the population density [10]. Such diversity creates considerable ambiguity about the traffic consumption at sub-cell scale and multiple 'high-resolution' snapshots could be matched to their aggregate counterparts. Inferring the precise traffic distribution is therefore hard.

**Drawing inspiration from image processing.** Spatio-temporal correlations we observe between mobile traffic patterns prompt us to represent these as tensors that highly resemble images (cross-spatial relations) or videos (cross-temporal relations). It becomes apparent that a problem similar to the one we tackle exists in the image processing field, where images with small number of pixels are enhanced to high-resolution. There, a super-resolution (SR) imaging approach mitigates the multiplicity of solutions by constraining the solution space through prior information [11]. This inspires us to employ image processing techniques to learn end-to-end relations between low- and high-resolution mobile traffic snapshots. We illustrate the similarity between the two SR problems in Fig. 1. Recent achievements in GPU based computing [12] have led to important results in image classification [13], while different neural network architectures have been successfully employed to learn complex relationships between low- and high-resolution images [14, 15]. Thus we recognise the potential of exploiting deep learning models to achieve reliable mobile traffic super-resolution (MTSR) and make the following **key contributions:**

(1) We propose a novel Generative Adversarial neural Network (GAN) architecture tailored to MTSR, to infer fine-grained mobile traffic patterns, from aggregate measurements collected by network probes. Specifically, in our design high-resolution traffic maps are obtained through a *generative model* that outputs approximations of the real traffic distribution. This is trained with a *discriminative model* that estimates the probability a sample snapshot comes from a fine-grained ground truth measurements set, rather than being produced by the generator.

(2) We construct the generator component of the GAN using an original deep zipper network (ZipNet) architecture. This upgrades a sophisticated ResNet model [16] with a set of additional 'skip-connections', without introducing extra parameters, while allowing gradients to backpropagate faster through the model in the training phrase. The ZipNet further introduces 3D upscaling blocks to jointly extract spatial and temporal features that exist in mobile traffic patterns. Our design accelerates training convergence and we demonstrate it outperforms the baseline Super-Resolution Convolutional Neural Network (SRCNN) [14] in terms of prediction accuracy.

(3) To stabilise the adversarial model training process, and prevent model collapse or non-convergence problems (common in GAN design), we introduce an empirical loss function. The intuition behind our choice is that the generator can be optimised and complexity reduced, if using a loss function for model training that is insensitive to model structure and hyper-parameters configuration.

(4) We propose a data processing and augmentation procedure to handle the insufficiency of training data and ensure the neural network model does not turn significantly over-fitted. Our approach crops the original city-wide mobile data 'snapshots' to smaller size windows and repeats this process with different offsets to generate extra data points from the original ones, thereby maximising the usage of data sets available for training.

(5) We conduct experiments with a publicly available real-world mobile traffic data set and demonstrate the proposed ZipNet (-GAN) precisely infer fine-grained mobile traffic distributions with up to 100× higher granularity as compared to standard probing, irrespective to the coverage and the position of the probes. Importantly, our solutions outperform existing traditional and deep-learning based interpolation methods, as we achieve up to 78% lower reconstruction errors, 40% higher fidelity of reconstructed traffic pattern, and improve the structural similarity by 36.4×.

We believe the proposed ZipNet(-GAN) techniques can be deployed at a gateway level to effectively reduce the complexity and enhance the quality of mobile traffic analysis, by simplifying the networking infrastructure, underpinning intelligent resource management, and overcoming service congestion in popular 'hot spots'.

## 2 PROBLEM FORMULATION

The objective of MTSR is to infer city-wide fine-grained mobile data traffic distributions, using sets of coarse-grained measurements collected by probes deployed at different locations. We formally define low-resolution traffic measurements as a spatio-temporal sequence of data points $\mathcal{M}^L = \{D_1^L, D_2^L, \ldots, D_T^L\}$, where $D_t^L$ is a snapshot at time $t$ of the mobile traffic consumption summarised over the entire coverage area and in practice partitioned into $V$ cells (possibly of different size), i.e. $D_t^L = \{l_t^1, \ldots, l_t^V\}$. Here $l_t^v$ represents the data traffic consumption in cell $v$ at time $t$.

We denote $\mathcal{M}^H = \{D_1^H, D_2^H, \ldots, D_T^H\}$ the high-resolution mobile traffic measurement counterparts (which are traditionally obtained via aggregation and post-processing), where $D_t^H$ is a mobile traffic consumption snapshot at time $t$ over $I$ sub-cells, i.e. $D_t^H = \{h_t^1, \ldots, h_t^I\}$. Here $h_t^i$ denotes the data traffic volume in sub-cell $i$ at time $t$. Dissimilar to $\mathcal{M}^L$, we work with sub-cells of the same size and shape, therefore $D_t^H$ points have the same measuring granularity. Note that both $\mathcal{M}^L$ and $\mathcal{M}^H$ measure the traffic consumption in the same area and for the same duration.

From a machine learning prospective, the MTSR problem is to infer the most likely current fine-grained mobile traffic consumption, given previous $S$ observations of coarse-grained measurements. Denoting this sequence $F_t^S = \{D_{t-S+1}^L,$
$\ldots, D_t^L\}$, MTSR solves the following optimisation problem:

$$\widetilde{D}_t^H := \arg\max_{D_t^H} p\left(D_t^H | F_t^S\right), \quad (1)$$

where $\widetilde{D}_t^H$ denotes the solution of the prediction. Both $D_t^H$ and $F_t^S$ are high-dimensional, since they represent different traffic measurements across a city. To precisely learn the complex correlation between $D_t^H$ and $F_t^S$, in this work we propose to use a Generative Adversarial Network (GAN), which will model the conditional distribution $p\left(D_t^H | F_t^S\right)$. As we will demonstrate, the key advantage of employing a GAN structure is that it will not only minimise the mean square error between predictions and ground truth, but also yield remarkable fidelity of the high-resolution inferences made.

## 3 PERFORMING MTSR VIA ZIPNET-GAN

In what follows we propose a deep-learning approach to tackle the MTSR problem using GANs. This is a novel unsupervised learning framework for generation of artificial data from real distributions through an adversarial training process [17]. In general, a GAN is composed of two neural network models, a generator $\mathcal{G}$ that learns the data distribution, and a discriminative model $\mathcal{D}$ that estimates the probability that a data sample came from the real training data rather than from the output of $\mathcal{G}$. We first give a brief overview of general GAN operation, then explain how we adapt this structure to the MTSR problem we aim to solve.

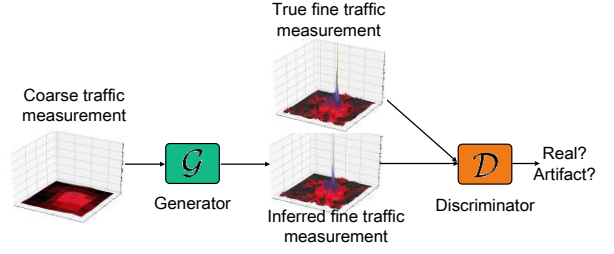

**Figure 2: GAN operating principle in MTSR problem. Only the generator is employed in prediction phase.**

### 3.1 Essential Background

As with all neural networks, the key to their performance is the training process that serves to configure their parameters. When training GANs, the generator $\mathcal{G}$ takes as input a noisy source (e.g. Gaussian or uniformly distributed) $z \sim P_n(z)$ and produces an output $\hat{x}$ that aims to follow a target unknown data distribution (e.g. pixels in images, voice samples, etc.). On the other hand, the discriminator $\mathcal{D}$ randomly picks data points generated by $\mathcal{G}$, i.e. $\hat{x} \sim \mathcal{G}(z)$, and others sampled from the target distribution $x \sim P_r(x)$, and is trained to maximise the probabilities that $\hat{x}$ is fake and $x$ is real. In contrast, $\mathcal{G}$ is trained to produce data whose distribution is as close as possible to $P_r(x)$, while maximising the probability that $\mathcal{D}$ makes mistakes. This is effectively a two-player game where each model is trained iteratively while fixing the other one. The joint objective is therefore to solve the following minimax problem [17]:

$$\min_{\mathcal{G}} \max_{\mathcal{D}} \mathbb{E}_{x \sim P_r(x)}[\log \mathcal{D}(x)] + \mathbb{E}_{z \sim P_n(z)}[\log(1 - \mathcal{D}(\mathcal{G}(z)))]. \quad (2)$$

Once trained, the generator $\mathcal{G}$ is able to produce artificial data samples from the target distribution given noisy input $z$.

In the case of our MTSR problem, the input of $\mathcal{G}$ is sampled from the distribution of coarse-grained measurements $p(F_t^S)$, instead of traditional Gaussian or uniform distributions. Our objective is to understand the relations between $D_t^H$ and $F_t^S$, i.e. modelling $p\left(D_t^H | F_t^S\right)$. $\mathcal{D}$ is trained to discriminate whether the data is a real fine-grained traffic measurement snapshot, or merely an artefact produced by $\mathcal{G}$. We summarise this principle in Fig. 2.

### 3.2 The ZipNet-GAN Architecture

Recall that our GAN is composed of a generator $\mathcal{G}$ that takes low-resolution measurements $F_t^S$ as input and reconstructs their high-resolution counterparts $D_t^H$, and a discriminator $\mathcal{D}$ whose task is relatively light (learning to discriminate samples and minimising the probability of making mistakes).

In general the generator has a complicated structure, which is required for the reconstruction of data points with unknown distributions. In our MTSR problem, we want to capture the complex correlations between $F_t^S$ and $D_t^H$. To this end, we propose Zipper Network (ZipNet), an original deep neural network architecture. This upgrades a ResNet model [16] with additional 'skip-connections', without introducing extra parameters, while allowing gradients to backpropagate faster through the model, and accelerate the training process. The overall ZipNet architecture is illustrated in Fig. 3. To exploit important historical traffic information, we introduce 3D upscaling blocks that extract the spatio-temporal features specific to mobile traffic patterns. The proposed ZipNet comprises three main components, namely:

• **3D upscaling blocks** consisting of a 3D deconvolutional layer [18], three 3D convolutional layers [19], a batch normalisation (BN) layer [20], and a Leaky ReLU (LReLU) activation layer [21]. 3D (de)convolutions are employed to capture spatial (2D) and temporal relations between traffic measurements. The deconvolution is a transposed convolutional layer widely employed for image upscaling. 3D convolutional layers enhance the model representability. BN layers normalise the output of each layer and are effective in training acceleration [20]. LReLUs improve the model non-linearity and their output is of the form:

$$\text{LReLu}(x) = \begin{cases} x, & x \geq 0; \\ \alpha x, & x < 0, \end{cases} \quad (3)$$

where $\alpha$ is a small positive constant (e.g. 0.1). The objective of such upscaling blocks is to up-sample the input $F_t^S$ into tensors that have the same spatial resolution as the desired output $D_t^h$. The number of upscaling blocks increases with the resolution of the input (from 1 to 3). **These 3D upscaling blocks are key to jointly extracting spatial and temporal features specific to mobile traffic.**

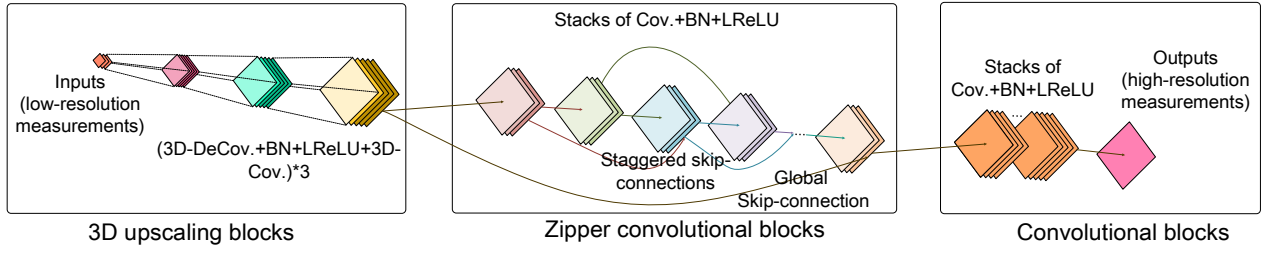

**Figure 3:** The deep Zipper Network (ZipNet) generator architecture, consisting of 3D upscaling, zipper convolutional, and standard convolutional blocks. Conv. and deconv. denote convolutional and deconvolutional layers.

• **Zipper convolutional blocks (core)**, which pass the output of the 3D upscaling blocks, through 24 convolutional layers, BN layers, and LReLU activations, with staggered and global skip connections. These block hierarchically extract different features and construct the input for the next convolutional component. We give further details about this block in the rest of this sub-section.

• **Convolutional blocks** that summarise the features distilled by the Zipper blocks and make the final prediction decisions. The block consists of three standard convolutional layers, BN layers, and LReLU activations, without skip connections. Each layer is configured with more feature maps as compared to the previous

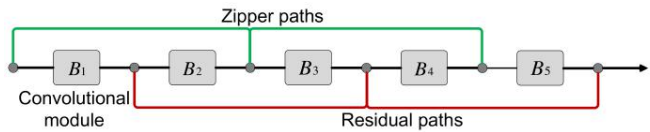

**Figure 4:** 5 Zipper convolutional blocks; each module $B$ includes a convolutional layer, a BN layer and a LReLU activation. Circular nodes represent additions.

block, in order to provide sufficient features for the mobile traffic final prediction.

The proposed ZipNet is a complex architecture that has over 50 layers. In general, accuracy grows with the depth (which is precisely what we want to achieve), but may also degrade rapidly, if the network becomes too deep. To overcome this problem, we use shortcut connections at different levels between the comprising blocks, as illustrated in Fig. 4 –this resembles a zipper, hence the naming. The structure can be regarded as an extension to residual networks (ResNets) previously used for image recognition [16], where the zipper connections significantly reduce the convergence rate and improve the model's accuracy, without introducing extra parameters and layers.

The fundamental module $B$ of the zipper convolutional block comprises a convolutional layer, a BN layer, and an activation function. Staggered skip connections link every two modules, and a global skip connection performs element-wise addition between input and output (see Fig. 3) to ensure fast backpropagation of the gradients. The skip connections also build an ensemble of networks with various depths, which has been proven to enhance the model's performance [22]. Further, replacing layer-wise connections (as e.g. in [23]) with light-weight staggered shortcuts reduces the computational cost. Compared to the original ResNets, extra zipper paths act as an additional set of residual paths, which reinforce the ensembling system and alleviate the performance degeneration problem introduced by deep architectures. The principle behind ensembling systems is collecting a group of models and voting on their output for prediction (e.g. using a random forest tool). Extra zipper paths increase the number of candidate models in the voting process, which improves the robustness of the architecture and contributes to superior performance.

Finally, the discriminator $\mathcal{D}$ accepts simultaneously the predictions made by the generator and fine-grained ground truth measurements, and minimises the probability of misjudgement. In our design $\mathcal{D}$ follows a simplified version of a VGG-net neural network architecture [24] routinely used for imaging applications, and consists of 6 convolutional blocks (convolutional layer + BN layer + LReLU), as illustrated in Fig. 5. The number of feature maps doubles every other layer, such that the final layer will have sufficient features for accurate decision making. The final layer employs a sigmiod activation function, which constrains the output to a probability range, i.e. (0, 1).

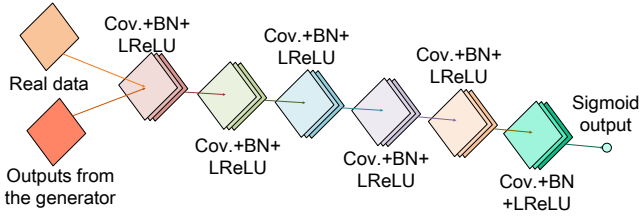

**Figure 5: Architecture of the discriminator $\mathcal{D}$ we use in ZipNet-GAN, based on the VGG-net structure [24].**

The remaining task in the design of the proposed neural network is to configure the loss functions employed by the generator and discriminator, which is key to the adjustment of the weights within the different layers.

### 3.3 Designing the Loss Functions

The MTSR is a supervised learning problem, hence directly deploying the traditional adversarial training process is inappropriate. This is because the data generated by $\mathcal{G}$ in traditional GANs is stochastically sampled from the approximated target distribution, whereas in MTSR we expect the output of the generator to follow as close as possible the distribution of real fine-grained measurements. That is, individual data points $\mathcal{G}(F_t^S)$ produces should be close to their corresponding ground truth. This requires to minimise *(i)* the divergence between the distributions of the real data $p(D_t^H)$ and the generated data $p(\mathcal{G}(F_t^S))$, and *(ii)* the Euclidean distance between individual predictions and their corresponding ground truth.

$L_2$ loss functions commonly used in neural networks to solve the optimisation problem given in (1) only fulfil the second objective above, but do not guarantee that the generator's output will follow the real data distribution. This may lead to failure to match the fidelity expected of high-resolution predictions [25]. Therefore we add another term to (1), which requires to design different loss functions, while changing the original problem to

$$\widetilde{D}_t^H = \arg\max_{D_t^H} p\left(D_t^H | F_t^S\right) \cdot \mathcal{D}(\mathcal{G}(F_t^S)), \quad (4)$$

where recall that $\mathcal{D}(\mathcal{G}(F_t^S))$ is the probability that $\mathcal{D}$ predicts whether data is sampled from the real distribution. Minimising this new term $\mathcal{D}(\mathcal{G}(F_t^S))$ aims to 'fool' the discriminator, so as to minimise the divergence between real and generated data distributions, i.e. objective *(i)* above, and remedy the aforementioned fidelity problem.

The discriminator $\mathcal{D}$ works as a binary classifier, which is trained to configure the discriminator's parameters $\Theta_{\mathcal{D}}$ by maximising the following loss function:

$$\hat{\mathcal{L}}(\Theta_{\mathcal{D}}) = \sum_{t=S}^{T} \log \mathcal{D}(D_t^h) + \log(1 - \mathcal{D}(\mathcal{G}(F_t^S))). \quad (5)$$

The more challenging part is the design of the loss function employed to train the generator. Following the common hypothesis used in regression problems, we assume that the prediction error $\epsilon$ follows a zero-mean Gaussian distribution, with diagonal covariance matrix $\sigma^2 \mathbf{I}$ [26], i.e.

$$\epsilon \sim N(\epsilon | \mathbf{0}, \sigma^2 \mathbf{I}). \quad (6)$$

Then a fine-grained data point $D_t^H$ can be approximated with respect to the corresponding prediction error as following a multivariate Gaussian distribution with a mean that depends on $F_t^S$ and a diagonal covariance matrix $\sigma^2 \mathbf{I}$. This allows us to rewrite the conditional distribution in (4) as

$$p\left(D_t^H | F_t^S\right) \sim N(D_t^H | \mathcal{G}(F_t^S), \sigma^2 \mathbf{I}). \quad (7)$$

Substituting (7) in (4), and adopting maximum likelihood estimation, the problem (4) is equivalent to minimising the following loss function, in order to configure the parameters $\Theta_{\mathcal{G}}$ of the generator:

$$\mathcal{L}(\Theta_{\mathcal{G}}) = \frac{1}{T} \sum_{t=S}^{T} \left[ ||D_t^H - \mathcal{G}(F_t^S)||^2 - 2\sigma^2 \log \mathcal{D}(\mathcal{G}(F_t^S)) \right]. \quad (8)$$

Recall that $\sigma^2$ is the variance of the Gaussian distribution of $\epsilon$, which can be considered as a trade-off weight between the mean square error (MSE) $||D_t^H - \mathcal{G}(F_t^S)||^2$ and the adversarial loss $\log \mathcal{D}(\mathcal{G}(F_t^S))$. The same loss function is used in [15] for image SR purposes, while the weight $\sigma^2$ is manually set.

However, **we find that the training process is highly sensitive to the configuration of $\sigma^2$**. Specifically, the loss function does not converge when $\sigma^2$ is large, while the discriminator rapidly reach an optimum if $\sigma^2$ is small, which in turn may lead to model collapse [27] and overall poor performance. To solve this problem, we propose an alternative loss function in which we replace the $\sigma^2$ term with the MSE. This yields

$$\hat{\mathcal{L}}(\Theta_{\mathcal{G}}) = \frac{1}{T} \sum_{t=S}^{T} (1 - 2 \log \mathcal{D}(\mathcal{G}(F_t^S))) \cdot ||D_t^H - \mathcal{G}(F_t^S)||^2. \quad (9)$$

In the above, the MSE term forces the predicted individual data points to be close to their corresponding ground truth, while the adversarial loss works to minimise the divergence between two distributions. Our experiments suggest that this new loss function significantly stabilises the training process, as the model never collapses and the process converges fast. In what follows, we detail the training procedure.

### 3.4 Training the ZipNet-GAN

To train the ZipNet-GAN we propose for solving the MTSR problem, we employ Algorithm 1, which takes a Stochastic Gradient Descent (SGD) approach and we explain the steps involved next. Recall that the purpose of training is to configure the parameters of the neural network components, $\Theta_{\mathcal{G}}$ and $\Theta_{\mathcal{D}}$. We work with the Adam optimiser [28], which yields faster convergence as compared to traditional SGD.

We train $\mathcal{G}$ and $\mathcal{D}$ iteratively, each time for $n_{\mathcal{G}}$ and $n_{\mathcal{D}}$ sub-epochs, by fixing the parameters of one and configuring the others', and vice-versa, until their loss functions converge (line 3). At every sub-epoch (lines 4, 9), $\mathcal{G}$ and $\mathcal{D}$ randomly sample $m$ low-/high-resolution traffic measurement pairs (lines 5, 10) and compute the gradients $g_{\mathcal{G}}$ and $g_{\mathcal{D}}$ (lines 6, 11) to be used in the optimisation (lines 7, 12).

The key to this training process is that $\mathcal{G}$ and $\mathcal{D}$ make progress synchronously. At early training stages, when $\mathcal{G}$ is poor, $\mathcal{D}$ can

**Algorithm 1** The GAN training algorithm for MTSR.

1: **Inputs:**
   Batch size $m$, low-/high-res traffic measurements $\mathcal{M}^L$ and $\mathcal{M}^H$, learning rate $\lambda$, generator and discriminator sub-epochs, $n_\mathcal{G}$ and $n_\mathcal{D}$.
2: **Initialise:**
   Generative and discriminative models, $\mathcal{G}$ and $\mathcal{D}$, parameterised by $\Theta_\mathcal{G}$ and $\Theta_\mathcal{D}$.
   Pre-trained $\mathcal{G}$ by minimising (10).
3: **while** $\Theta_\mathcal{G}$ and $\Theta_\mathcal{D}$ not converge **do**
4:    **for** $e_\mathcal{D} = 1$ to $n_\mathcal{D}$ **do**
5:       Sample low-/high-res traffic measurement pairs $\{F_t^S, D_t^H\}_{t=1}^m, F_t^S \in \mathcal{M}^L, D_t^H \in \mathcal{M}^H$.
6:       $g_\mathcal{D} \leftarrow \Delta_{\Theta_\mathcal{D}}[\frac{1}{m}\sum_{i=1}^m \log \mathcal{D}(D_t^H) +$
   $+\frac{1}{m}\sum_{i=1}^m \log(1 - \mathcal{D}(\mathcal{G}(F_t^S)))]$.
7:       $\Theta_\mathcal{D} \leftarrow \Theta_\mathcal{D} + \lambda \cdot \text{Adam}(\Theta_\mathcal{D}, g_\mathcal{D})$.
8:    **end for**
9:    **for** $e_\mathcal{G} = 1$ to $n_\mathcal{G}$ **do**
10:     Sample low-/high-res traffic measurement pairs $\{F_t^S, D_t^H\}_{t=1}^m, F_t^S \in \mathcal{M}^L, D_t^H \in \mathcal{M}^H$.
11:     $g_\mathcal{G} \leftarrow \Delta_{\Theta_\mathcal{G}} \frac{1}{m}\sum_{t=1}^m (1 - 2\log \mathcal{D}(\mathcal{G}(F_t^S))) \cdot$
   $\cdot \|D_t^H - \mathcal{G}(F_t^S)\|^2$.
12:     $\Theta_\mathcal{G} \leftarrow \Theta_\mathcal{G} - \lambda \cdot \text{Adam}(\Theta_\mathcal{G}, g_\mathcal{G})$.
13:    **end for**
14: **end while**

reject samples with high confidence as they are more likely to differ from the real data distribution. An ideal $\mathcal{D}$ can always find a decision boundary that perfectly separates true and generated data points, as long as the overlapping measure support set of these two distributions is null. This is highly likely in the beginning and can compromise the training of $\mathcal{G}$. To overcome this issue, we initialise the generator by minimising the following MSE until convergence

$$\text{MSE}(\Theta_\mathcal{G}) = \frac{1}{T}\sum_{t=S}^T \|D_t^H - \mathcal{G}(F_t^S)\|^2. \qquad (10)$$

Note that at this stage the initialised $\mathcal{G}$ (line 2) could be directly deployed for the MTSR purpose, since the MSE based initialisation is equivalent to solving (1). However, this only minimises the point-wise Euclidean distance, and the further training steps in Algorithm 1 are still required to ensure the predictor does not lose accuracy. In our experiments, we set $n_\mathcal{G}$ and $n_\mathcal{D}$ to 1, and learning rate $\lambda$ to 0.0001.

Next we discuss the processing and augmentation of the data set we employ to train and evaluate the performance of the proposed ZipNet(-GAN).

## 4 DATA PROCESSING & AUGMENTATION

To train and evaluate the ZipNet(-GAN) architecture, we conduct experiments with a publicly available real-world mobile traffic data set released through Telecom Italia's Big Data Challenge [29]. This contains network measurements in terms of total cellular traffic volume observed over 10-minute intervals in Milan, between 1 Nov 2013 and 1 Jan 2014 (2 months). This was obtained by combining call detail records (CDR) that were generated upon user interactions with base stations, namely each time a user started/ended an Internet connection, or a user consumed more than 5 MB. The coverage area is partitioned into $100 \times 100$ squares of $0.055\text{km}^2$ (i.e. $235\text{m} \times 235\text{m}$). We show snapshots of the traffic's spatial distribution for both off-peak and peak times in Fig. 6.

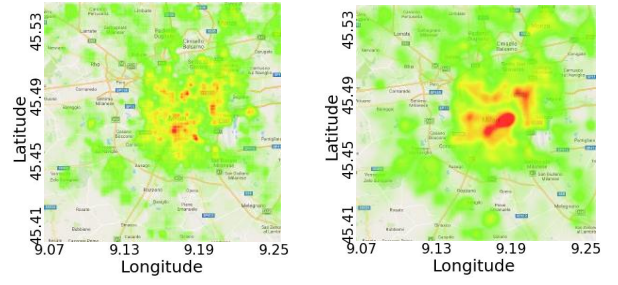

**Figure 6: Spatial distribution of mobile data traffic consumption in Milan during off-peak (left) and peak (right) times. Traffic consumption per 10-minute interval varies between 20 MB (green) to 5,496 MB (red). Figure best viewed in colour.**

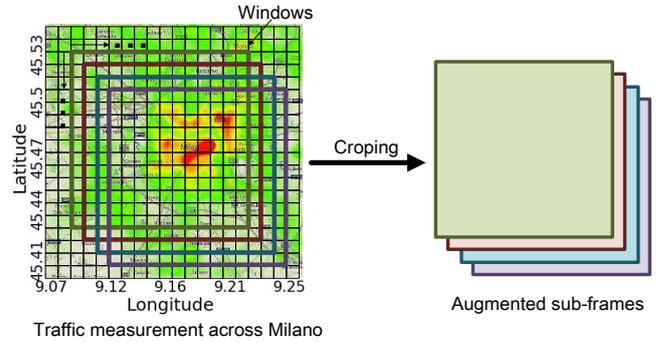

**Figure 7: Illustration of the data augmentation technique we propose to enable precise training in view of performing MTSR. Figure best viewed in colour.**

Overall the data set includes $T = 8,928$ snapshots of fine-grained mobile traffic measurements. This number appears sufficient when using traditional machine learning tools, though it is grossly insufficient for the purpose of training a complex neural network, as in our case. This is due to the highly dimensional parameter set, which can turn the model over-fitting if improperly configured through training with small data sets. To overcome this problem, we augment the Milan data set by cropping each snapshot to multiple 'windows' of smaller size, each with different offsets (1 cell increment in every dimension). We illustrate this augmentation process in Fig. 7.

We work with 'sub-frames' of $80 \times 80$ sub-cells, thereby producing 441 new data points from every snapshot. Note that the initial size of model's prediction matrix is also $80 \times 80$ and we employ a moving average filter to construct the final predictions across the original grid (i.e. $100 \times 100$).

## 5 EXPERIMENTS

In this section we first describe briefly the implementation of the proposed ZipNet(-GAN), then evaluate its performance in terms of prediction accuracy (measured as Normalised Mean Root Square Error – NMRSE), fidelity of the inferences made (measured through Peak Signal-to-Noise Ratio – PSNR), and similarity between predictions made and ground truth measurements (structural similarity index – SSIM). We compare the performance of ZipNet(-GAN) with that of Uniform and Bicubic interpolation techniques [30], Sparse Coding method (SC) [31], Adjusted Anchored Neighbouring Regression (A+) [32], and Super-Resolution Convolutional Neural Network (SRCNN) previously used in imaging applications [14].

### 5.1 Implementation

We implement the proposed ZipNet(-GAN), the SRCNN, the Uniform and Bicubic interpolation methods using the open-source Python libraries TensorFlow [12] and TensorLayer [33].[1] We train the models using a GPU cluster comprising 19 nodes, each equipped with 1-2 NVIDIA TITAN X and Tesla K40M computing accelerators (3584 and respectively 2280 cores). To evaluate their prediction performance, we use one machine with one NVIDIA GeForce GTX 970M GPU for acceleration (1280 cores). SC and A+ super-resolution techniques are implemented using Matlab. In what follows, we detail four MTSR instances with different granularity, which we use for comparison of the different SR methods.

### 5.2 Different MTSR Granularity

In practice measurement probes are deployed at different locations (e.g. depending on user density) and often have different coverage in terms of number of cells. We approximate the coverage of a probe by a square area consisting of $r_f = n_f \times n_f$ sub-cells, and refer to $n_f$ as an upscaling factor. The smaller the $n_f$, the higher the granularity of measurement. Given the heterogeneous nature of cellular deployments, we construct four different MTSR instances with different granularity, as summarised in Table 1.

| Instance | Configuration | $n_f$ (Avg.) | $r_f$ (Avg.) |
|---|---|---|---|
| Up-2 | Probes cover $2 \times 2$ sub-cells | 2 | 4 |
| Up-4 | Probes cover $4 \times 4$ sub-cells | 4 | 16 |
| Up-10 | Probes cover $10 \times 10$ sub-cells | 10 | 100 |
| Mixture | 7% of probes cover $10 \times 10$ sub-cells, 44% cover $4 \times 4$, and 49% cover $2 \times 2$ sub-cells. | 4 | 16 |

Table 1: Configuration of MTSR instances with different granularity considered.

The first three instances assume that all probes are uniformly distributed in Milan and have the same coverage, i.e. 4, 16, and 100 sub-cells respectively. The fourth instance corresponds to a mixture of probes, each with different coverage, as we illustrate in Fig. 8. Specifically, we consider three types of probes covering 2×2, 4×4, and respectively 10×10 sub-cells. In this instance, more probes serve the city centre (red area), each with smaller coverage and thus finer granularity. Fewer probes are assumed in surrounding areas, covering larger regions (the yellow and green areas). We compute aggregate measurements collected by such dissimilar probes, each aggregate corresponding to the sum over the covered cells. We then project these points onto a square that becomes the input to our model, as shown on the right in Fig. 8. Therefore, for each traffic snapshot we can obtain 400 measurement points that can be viewed as a 2D square (right side of Fig. 8), which we use for training and performing inferences. Note that although the average $n_f$ of the mixture instance is 4, its structure differs significantly from that of the 'up-4' instance. Our evaluation will reveal that such differences will affect the accuracy of the MTSR outcome.

[1] The source-code of our implementation is available at https://github.com/vyokky/ZipNet-GAN-traffic-inference.

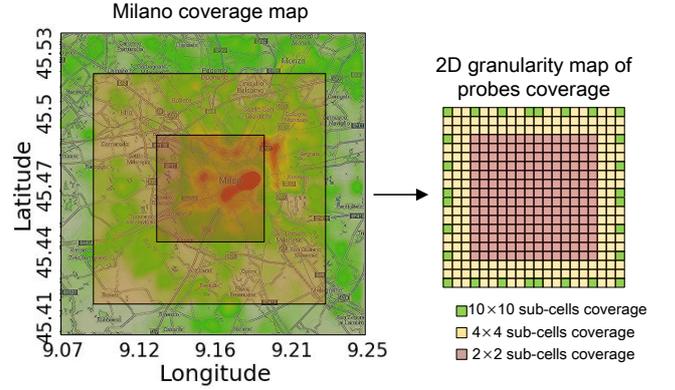

Figure 8: Cellular deployment map for Milan served by a mixture of probes with different coverage (left), and the corresponding measurement granularity projected onto a 2D map (right). Best viewed in colour.

We train all deep learning models with data collected between 1 Nov 2013 and 10 Dec 2013 (40 days), validate their configuration with measurements observed over the following 10 days, and finally evaluate their performance using the averages aggregated by the different probes (different granularity) between 20–30 Dec 2013. Prior to training, all data is normalised by subtraction of the mean and division by the standard deviation. The coarse-grained input measurement data (i.e. $\mathcal{M}^L$) is generated by averaging the traffic consumption over cells' coverage.

### 5.3 Performance Evaluation

**Evaluation Metrics:** We quantify the accuracy of the proposed ZipNet(-GAN), comparing against that of existing super-resolution methods, in terms of Normalised Root Mean Square Error (NRMSE), Peak Signal-to-Noise Ratio (PSNR), and Structural Similarity Index (SSIM).

NRMSE is computed over $I$ sub-cells at time $t$ as follows:

$$\text{NRMSE} = \frac{1}{\overline{D}_t^H} \sqrt{\sum_{i=1}^{I} \frac{(\widetilde{h}_t^i - h_t^i)^2}{I}}, \qquad (11)$$

where $\widetilde{h}_t^i$ is the inferred high-resolution traffic consumption in a sub-cell $i$, $h_t^i$ is the corresponding ground truth value, and $\overline{D}_t^H$ is the ground truth mean. Note that NRMSE is frequently used in bioinformatics, experimental psychology, and other fields, for comparing data sets or models with different scales [34]. This is effectively the mean square deviation (the difference between predictions and ground truth) normalised by the mean value of ground truth samples. Low NRMSE values indicate less residual variance when the data sets subject to comparison may have different scales.

To understand the accuracy of the different MTSR techniques from different perspectives, we also compute the PSNR and SSIM, which are typically used to evaluate the similarity between images. PSNR is a common measure of quality of image reconstruction (following compression), and has the following expression:

$$\text{PSNR} = 20 \log(\max_i h_i) - 10 \log \frac{1}{I} \sum_{i=1}^{I} (h_t^i - \widetilde{h}_t^i)^2, \quad (12)$$

where $(\max_i h_i)$ is the highest traffic volume observed in one cell (in our case 5,496MB). SSIM is commonly used as an estimate of the perceived quality of images and video [35], and is computed with

$$\text{SSIM} = \frac{\left(2\overline{D}_t^H \cdot \overline{\widetilde{D}_t^H} + c_1\right)\left(2 \operatorname{cov}(D_t^H, \widetilde{D}_t^H) + c_2\right)}{\left((\overline{D}_t^H)^2 \cdot (\overline{\widetilde{D}_t^H})^2 + c_1\right)\left(\operatorname{var}(D_t^H)\operatorname{var}(\widetilde{D}_t^H) + c_2\right)}, \quad (13)$$

where $\overline{\widetilde{D}_t^H}$ is the average of the predictions, cov is the covariance between predictions and ground-truth, var denotes the variance of each data set, while $c_1$ and $c_2$ are parameters that stabilise the division with weak denominators. Unlike NRMSE, higher PSNR and SSIM values usually reflect greater similarity between ground truth and predictions.

**Techniques for comparison:** We summarise the performance of the proposed ZipNet(-GAN) against a range of existing interpolation or image super resolution techniques, including Uniform interpolation, Bicubic interpolation, Sparse Coding based methods (SC) [31], Adjusted Anchored Neighbouring Regression (A+) [32], and Super-Resolution Convolutional Neural Network (SRCNN) [14]. Uniform interpolation is routinely used by operators and assumes that the data traffic volume is uniformly distributed across cells [8]. The Bicubic interpolation is a popular non-parametric tool frequently used to enhance the resolution of images [30]. SC and A+ are commonly used as benchmarks in image super-resolution evaluation. Finally, SRCNN is a benchmark deep learning architectures that comprises three convolutional layers. We note that in our evaluation ZipNet is a simplified version of the ZipNet-GAN, which is purely trained with the mean square error (10), without the help of the discriminator.

**Assessing the quality of inferences:** We summarise the performance of the proposed ZipNet(-GAN) and that of existing SR techniques in terms of the aforementioned metrics in Fig. 9, where we use different bars for each of the four MTSR instances considered for inference (2, 4, and 10 upscaling, and upscaling mixture). The bars correspond to averages for inferences made over 10 days (i.e. 1440 snapshots).

Observe that **ZipNet-GAN achieves the best performance for all MTSR instances**, outperforming traditional super resolution schemes. In particular, although SC and A+ work well in image SR, their performance is inferior to that of simple Uniform and Bicubic interpolation techniques when performing MTSR. Our intuition is that the mobile traffic data has substantially different spatial structure and scale, as compared to imaging data. Therefore traditional SR approaches are unable to capture accurately the relations between low- and high- resolution traffic 'frames'. On the other hand, SRCNN works acceptably with low upscaling MTSR instances (i.e. 'up-4' and 'up-2'), but performs poorly when processing the 'up-10' instances. Unlike Uniform, A+, and Bicubic, the proposed ZipNet-GAN achieves an up to 65%, 76% and respectively 78% smaller NRMSE. The ZipNet-GAN further attains the highest PSNR and SSIM among all approaches, namely up to 40% higher PSNR and a remarkable 36.4× higher SSIM. Although perhaps more subtle to observe, the ZipNet-GAN is more accurate than the ZipNet (i.e. without employing a discriminator), which confirms the effectiveness of the GAN approach.

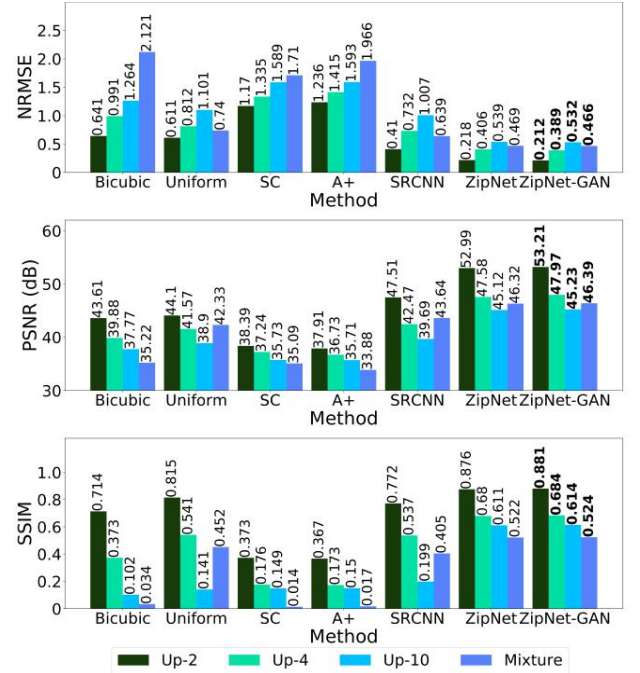

Figure 9: Comparison of inference accuracy of the proposed ZipNet(-GAN) and existing SR techniques in terms of NRMSE (above), PSNR (middle), SSIM (below). Four upscaling instances (different colour bars). Experimental results with the Milan data set [29].

Taking a closer look at Fig. 9, observe that the prediction accuracy of all approaches drops as the upscaling factor $n_f$ grows. This is indeed reasonable, since a larger $n_f$ corresponds to a greater degree of aggregation and thus detail information loss, which will pose more uncertainty to the models. Further, although the 'up-4' and mixture instances have the same average upscaling factor, the proposed ZipNet(-GAN) operate somewhat better with the former. Our intuition is that the mixture instance distorts the spatial correlation of the original measurements, as shown in Fig. 8. However, given the subtle performance differences, MTSR with probes of dissimilar coverage and granularity remains feasible with the proposed ZipNet-GAN.

We now delve deeper into the performance of all methods consider for MSTR and examine the behaviour of each with individual snapshots. To this end, Figs. 10 and 11 provide additional perspectives on the value of employing the proposed ZipNet(-GAN) to infer mobile traffic consumption with fine granularity. Each figure shows snapshots of the predictions made by all approaches, when working with the 'up-10' (Fig. 10) and the mixture MTSR (Fig. 11) instances. In particular, in Fig. 10 we observe that ZipNet(-GAN) deliver remarkably accurate predictions with a 99% reduction in term of measurement points required, as the texture and details are almost perfectly recovered. In contrast, the Uniform, Bicubic, SC, A+ and SRCNN techniques, although improve the resolution of data traffic 'snapshots', lose significant details and deviate considerable from the ground truth measurements (upper left corner).

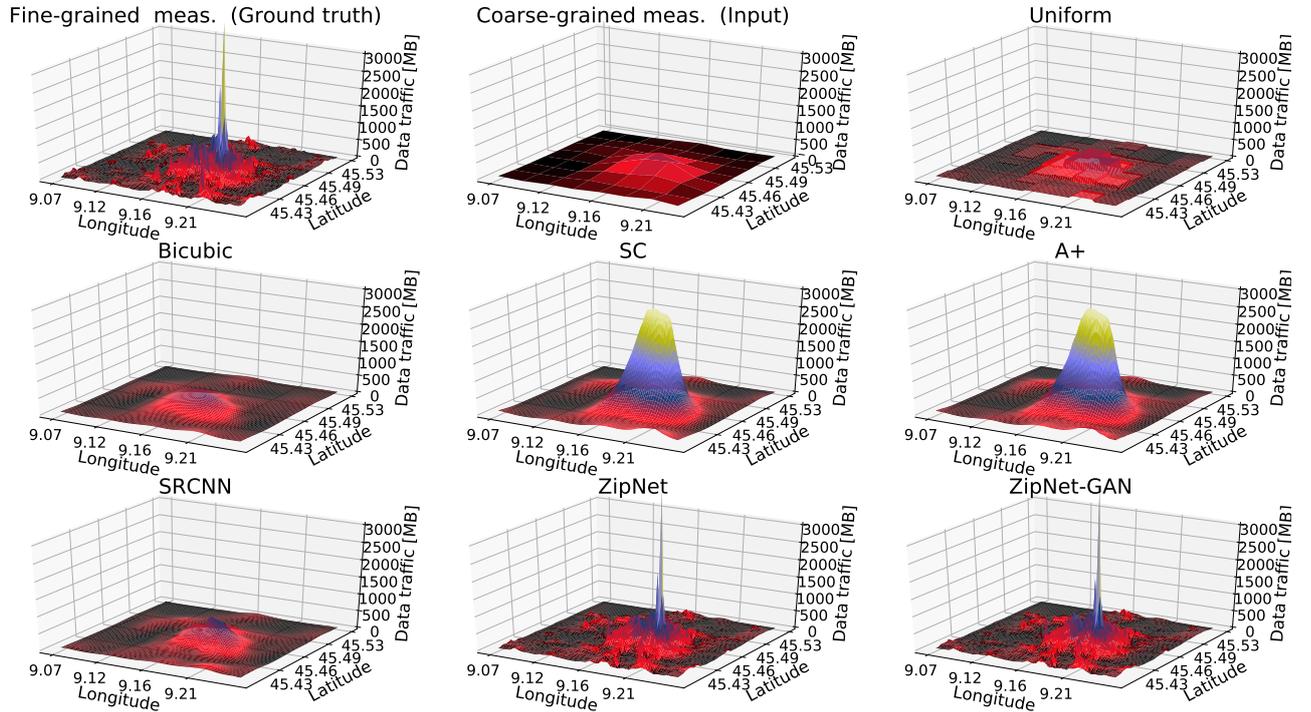

**Figure 10: Snapshots of ground truth, input data, and the *'up-10' MTSR instance* predicted by the proposed ZipNet (-GAN), existing interpolation methods and image SR techniques, using data collected in Milan on 21$^{st}$ Dec 2013.**

Turning attention to Fig. 11, observe that the input exhibits some spatial distortion (top centre plot), as the probes aggregating measurements have different coverage and locations. Despite this, the ZipNet(-GAN) still capture well spatial correlations and continues to perform very accurately (two plots in the bottom right corner). In contrast, the Uniform and Bicubic interpolations, although capture some spatial distribution mobile traffic profiles, significantly under-estimate the traffic volume in the city centre. In addition, the predictions made by SC and A+ exhibit significant distortions and yield inferior performance, demonstrating their image SR capabilities cannot be mapped directly to MTSR tasks. Lastly, the SRCNN approach, which employs deep learning, works acceptably in areas with low traffic intensity, but largely underestimates the traffic volume in the city centre.

### 5.4 The Benefits of GAN

In image SR, GAN architectures improve the fidelity of high-resolution output, making the image more photo-realistic. Here we show the GAN plays a similar role in MTSR. To this end, in Fig. 12 we present zoom snapshots of the predictions made by ZipNet and ZipNet-GAN. These offer a clear visual perception of the inference improvements of GAN in central parts of the city. Indeed, observe that including a GAN in the architecture improves the prediction fidelity, as it minimises the divergence between real and predicted data distributions, although this does not necessarily enhance overall accuracy. Note that the additional accuracy does not come at the cost of increased complexity, since the adversarial training is fast in terms of convergence and the discriminator will be abandoned in the inference phase.

### 5.5 Robustness to Abnormal Traffic

To evaluate the robustness of our solution in the presence of traffic anomalies, we artificially add such traffic patterns to the test data set and investigate the behaviour of our proposal. Specifically, we introduce abrupt traffic demands in suburban areas, which can be regarded as occurrences of social events (e.g. concert, football match, etc.), as seen in the bottom left corner of the second sub-plot in Fig. 13. Although such anomalies do not occur in the training set, the proposed ZipNet-GAN still successfully identifies the locations of abnormal traffic, given averaged and smoothed inputs (first sub-plot). This implies that, to some extent, our proposal can perform as an anomaly detector operating only with coarse measurements.

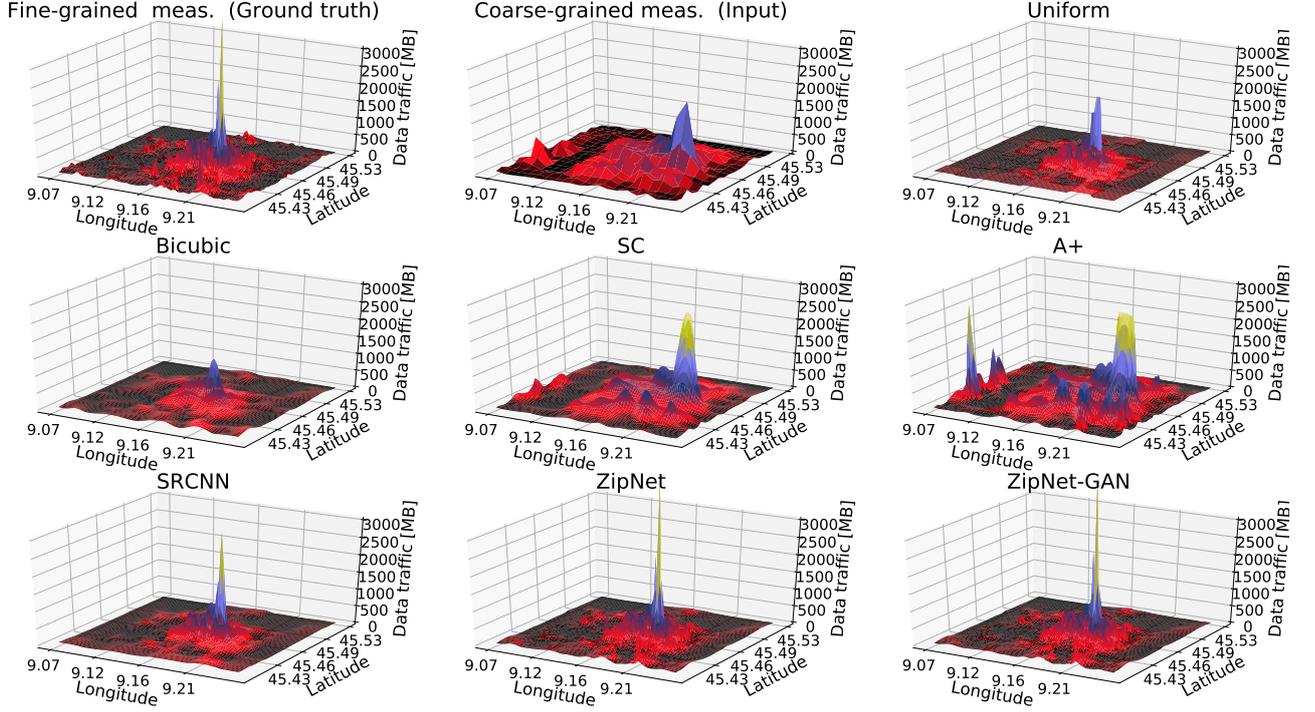

Figure 11: Snapshots of ground truth, input data, and the *mixture MTSR instance* predicted by the proposed ZipNet(-GAN), existing interpolation and image SR techniques, using data collected in Milan on 21$^{st}$ Dec 2013.

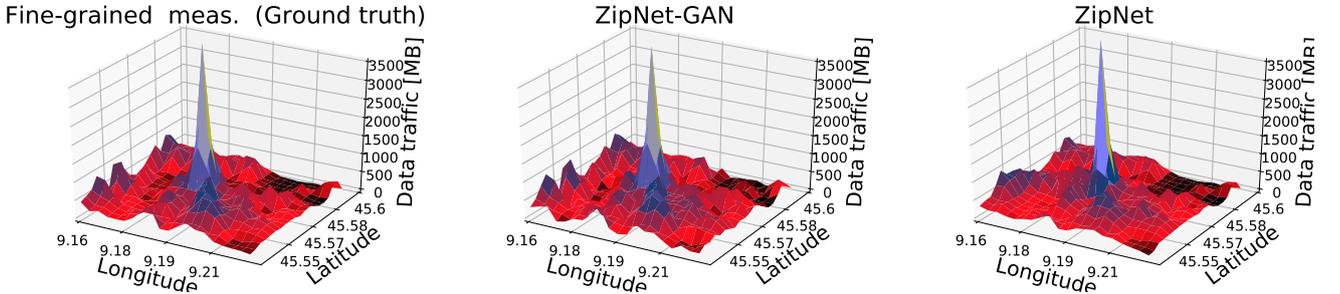

Figure 12: Zoom snapshots of ground truth and fine-grained traffic consumption predicted by the proposed ZipNet(-GAN) in an up-10 instance in central Milan.

## 5.6 Impact of Cross-Temporal Relations

We conclude our experiments by examining the impact cross-temporal correlations between traffic measurements provided as input have on the ZipNet-GAN architecture we propose. To this end, we *(i)* compare the MTSR accuracy with input of different temporal lengths $S$, and *(ii)* compute the absolute value of first-order derivatives of the loss function employed over input. The magnitude of these gradients are a good approximation of the sensitiveness of final prediction decisions to changes of the input [36].

**NRMSE with different length inputs:** We feed the proposed ZipNet-GAN with input of temporal length $S \in \{1, 3, 6\}$ snapshots, and illustrate in Fig. 14 the NRMSE attained with the three homogeneous MTSR instances considered. Observe that the prediction error drops with the increase of the number of snapshot we provide to our model in all instances, which indicates that earlier observations provide valuable insights toward inferring real fine-grained data traffic consumption. Additionally, the historical measurements play a more significant role with the increase of $n_f$ – in the 'up-10' instance the error between predictions made with $S = 1$ and $S = 6$ sequence lengths is much larger than in the same case for the other instances. This brings important implications to operators assessing the trade-offs between the length of inputs (which affects model complexity) and prediction accuracy.

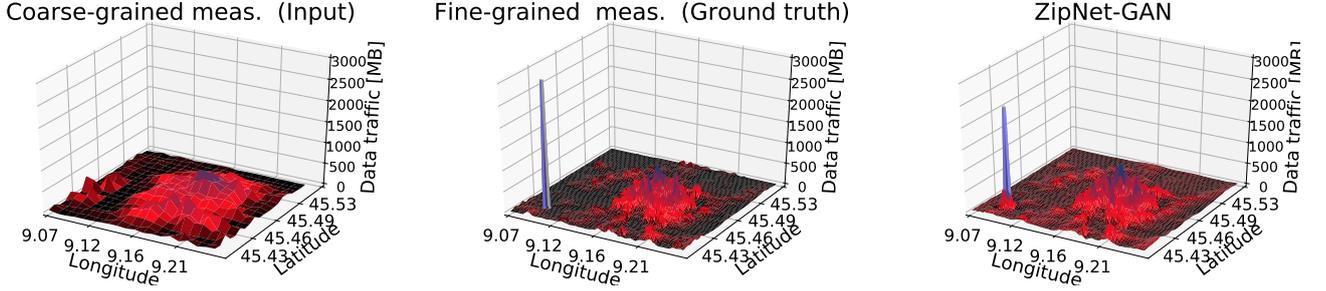

Figure 13: Behaviour in the presence of anomalies: snapshots of coarse-grained measurements fed as input (left), ground truth (middle) and fine-grained traffic patterns predicted by the proposed ZipNet-GAN (right) in a mixture instance.

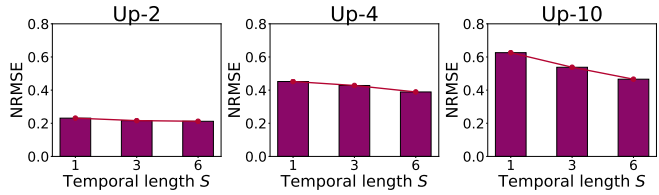

Figure 14: NRMSE comparison for three MTSR instances, with different temporal length $S$ of the input.

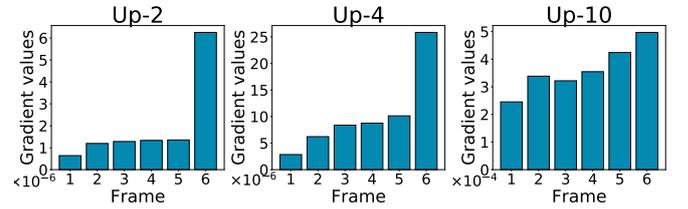

Figure 15: Mean magnitude of the gradient of the loss function $L(F_t^S)$ over inputs $F_t^S$, with different MTSR instances. Averages computed over the entire network across all test data.

**Magnitudes of gradients:** The impact of the input's temporal dimension can also be evaluated from the gradients' perspective. The loss function L is essentially a complex non-linear of the input $F_t^S$. However, this can be approximated by the first-order term of its Taylor expansion, i.e.

$$L(F_t^S) \approx w(F_t^S)^T \cdot F_t^S + b, \quad (14)$$

where $w(F_t^S)$ is the gradient of the loss function $L(F_t^S)$ over input $F_t^S$ and $b$ is a bias. Computing the absolute value of the gradient $w(F_t^S)^T$, i.e.

$$\frac{\partial L(F_t^S)}{\partial F_t^S} = \frac{\partial}{\partial F_t^S}\left[||D_t^H - \mathcal{G}(F_t^S)||^2(1 - 2\log \mathcal{D}(\mathcal{G}(F_t^S)))\right],$$

should give insights into the number of temporal steps required for accurate prediction with different MTSR instances. Therefore in Fig. 15 we plot the average magnitude of the gradient of the loss function over all inputs $F_t^S$, for all three homogeneous MTSR instances. Observe that the most recent 'frame' (i.e. frame 6) yields the largest gradient for all instances, as we expect. This means that the current measurement 'snapshot' provides the most valuable information for the model to reconstruct the fine-grained counterpart. Further, the contribution of historical measurements (i.e. frames 1 to 5) increases with the upscaling factor (from 2 to 10), which suggests that historical information becomes more significant when less spatial information is available, which is consistent with the insight gained earlier.

We conclude that, by exploiting the potential of deep neural networks and generative adversarial networks, the proposed ZipNet-GAN scheme can infer with high accuracy fine-grained mobile traffic patterns, and substantially outperforms traditional interpolation methods and image SR techniques, as it attains up to 78% lower prediction errors (NRMSE) and 36.4× higher structural similarity (SSIM). By employing a high performance GPU, training ZipNet-GAN takes 2-3 days, which we believe is negligible, considering the long-term costs of post-processing alternatives. Given the reasonable training time, ZipNet-GAN can be easily ported to different urban layouts after retraining for the target regions.

## 6 DISCUSSION

Obtaining city-scale fine-grained mobile traffic maps from coarse aggregates collected by probes brings important benefits to a range of applications critical to the urban realm. Namely,

- **Agile network resource management:** our approach enables infrastructure simplification and reduces OpEx. This is because base station controllers (BSC) are shared by increasing numbers of BSs, and already handle radio quality measurements and handovers. Adding further measurement capabilities, co-ordination of triangulation, and real-time processing of multi-modal measurements, poses scalability concerns, which may eventually impact on end user QoE. The proposed ZipNet-GAN shifts the burden to a single dedicated cluster that only requires to be trained infrequently and can perform coarse- to fine- grained measurement transformations, fed by summary statistics already available at selected probes.
- **Events localisation & response:** popular social events (e.g. concerts, fairs, sports, etc.) or emergency situations (fires, riots, or terrorist acts) exhibit high spatial similarity in terms of mobile data traffic consumption, which can serve as an

- indicator of density/size of the crowds involved. Accurate and timely localisation of such instances through inference of data traffic can help overcome network congestion in hot spots, but also appropriately provision policing and emergency services.
- **Context-based business:** accurate MTSR can also aid small businesses that can offer promotional products depending on spontaneous circumstances, e.g. families gathering in the park on a sunny day. Likewise university PR services can disseminate relevant information during highly-attended open/graduation days.

Therefore, we argue ZipNet-GAN is not limited to network infrastructure related tasks, e.g. precision traffic engineering or anomalous traffic detection [37], but could also serve civil applications such as events planning or even transportation. Importantly, once trained the proposed technique can continuously perform inferences on live streams, unlike post-processing approaches that only work off-line and thus have limited utility.

## 7 RELATED WORK

Although precise understanding of mobile data traffic distributions is essential for ISPs, research into accurate inference of fine-grained patterns is sparse, while the mobile traffic super-resolution problem we tackle in this work is unexplored.

**Traffic/Environment Measurement & Analysis:** Naboulsi *et al.* survey existing mobile traffic analysis practices, emphasising that cellular traffic monitoring relies on various probe types deployed across cities [5]. This includes RNC probes that record fine-grained state changes at each mobile device, but introduce substantial computational and storage costs; and PGW probes with larger geographical coverage that monitor mobile traffic aggregates with less overhead, but only offer approximate positioning information. None of these are well suited to accurate inference of fine-grained traffic patterns. Mobile traffic collected by such equipment has been analysed for different purposes. Furno *et al.* jointly classify the network-wide mobile traffic demand by spatial and temporal dimensions, to serve network activity profiling and land usage detection tasks [4]. Wang *et al.* extract and model traffic patterns in Shanghai to deliver insights into mobile traffic demand in large cities [3]. Their analysis is extended to forecast the traffic consumed across a city, using Autoregressive Integrated Moving Average (ARIMA) [2].

Roughan *et al.* infer Internet traffic matrices from data sets subject to severe measurement loss, using spatio-temporal compressive sensing (CS), and demonstrate superior performance over other interpolation methods [37]. Compressive sensing has also been employed to reconstruct sensory data collected in Wireless Sensor Networks, while exploiting spatial similarity and temporal stability of lossy/noisy measurements [38]. However, we recognise several reasons for which CS cannot be directly adopted for MTSR purposes. In particular, CS algorithms *(i)* require measurement matrices satisfying the Restricted Isometry Property, which may not be feasible with standard deterministic sampling constrained by the types and geographical distribution of probes; *(ii)* will not work with the highly non-linear relationships between coarse- and fine-grained samples, as they expect linear relationships between sparse traffic and inference matrices.

A crowdsourced based approach to urban sensing is introduced in [9], where city-scale sensor data is regarded as the output of a lossy 'urban camera', and temporal-correlation and collective decomposition are employed to infer air quality. Liu *et al.* define a metric to measure the quality of urban sensing, subsequently studying the relationship between sensing resolution and the number of smartphones/vehicles participating in such applications [39]. Compressive sensing and crowdsourcing measurement problems are different to the MTSR task we tackle in this work, since *in our case measurements are not subject to loss, noise, or incompleteness, but instead are complete aggregates that lack granularity.*

**Image Super-Resolution:** The super-resolution is becoming increasingly popular in the image processing field to reconstruct high-resolution images from compressed versions. SR techniques include early non-parametric methods such as Nearest, Bicubic and Lanczos interpolation [40], which unfortunately return output with overly smooth texture and distorted edges. Sparse coding [31] and neighbourhood regression methods [32, 41] overcome this issues as they achieve better representation and can correctly recover images from downsampled counterparts. More recently deep learning advances led to new major improvements in image SR performance. In particular, Chao *et al.* design a Super-Resolution Convolutional Neural Network (SRCNN) to learn the correlation between the low-/high-resolution images [14], while Generative Adversarial Network (GAN) architectures for accurate image SR are introduced in [15, 25]. Experiments demonstrate these can attain high mean opinion scores (MOS) [25] or high SSIM with large upscaling factor [15], when performing image SR. To our knowledge, the *ZipNet-GAN introduced in this paper is the first technique that tackles mobile traffic SR to make accurate inferences of fine-grained traffic consumption from coarse measurements.*

## 8 CONCLUSIONS

Precision large-scale mobile traffic analytics is increasingly vital for operators, to provision network resources agilely and keep up with end-user application requirements. Extracting fine-grained mobile traffic patterns is however complex and expensive, as this relies on specialised equipment, substantial storage capabilities, and intensive post-processing. In this paper we proposed a mobile traffic super resolution (MTSR) technique that overcomes these issues and infers mobile traffic consumption with fine granularity, given only limited coarse-grained measurement collected by probes. Our solution consists of a Generative Adversarial Network (GAN) architecture combined with an original deep Zipper Network (ZipNet) inspired by image processing tools. The proposed ZipNet-GAN is tailored to the specifics of the mobile networking domain and achieves reliable network-wide MTSR. Specifically, experiments we conduct with a real-world mobile traffic data set collected in Milan, demonstrate the proposed schemes predict narrowly localised traffic consumption, while improving measurement granularity by up to 100×, irrespective of the position and coverage of the probes. The ZipNet(-GAN) reduce the prediction error (NRMSE) of existing interpolation and super-resolution approaches by 78%, and achieve up to 40% higher fidelity (PSNR) and 36.4× greater structural similarity (SSIM).

## Acknowledgements

We thank Marco Fiore for his valuable suggestions that helped improving the quality of this paper. We also thank the anonymous reviewers for their constructive feedback and we are grateful for the guidance of our shepherd, Aruna Balasubramanian.

## REFERENCES

[1] Cisco, "Cisco Visual Networking Index: Global Mobile Data Traffic Forecast Update, 2016-2021 White Paper," Februrary 2017.

[2] F. Xu, Y. Lin, J. Huang, D. Wu, H. Shi, J. Song, and Y. Li, "Big data driven mobile traffic understanding and forecasting: A time series approach," *IEEE Transactions on Services Computing*, vol. 9, no. 5, pp. 796–805, Sept 2016.

[3] H. Wang, F. Xu, Y. Li, P. Zhang, and D. Jin, "Understanding mobile traffic patterns of large scale cellular towers in urban environment," in *Proc. ACM IMC*, 2015, pp. 225–238.

[4] A. Furno, M. Fiore, and R. Stanica, "Joint spatial and temporal classification of mobile traffic demands," in *Proc. IEEE INFOCOM*, 2017.

[5] D. Naboulsi, M. Fiore, S. Ribot, and R. Stanica, "Large-scale mobile traffic analysis: a survey," *IEEE Communications Surveys & Tutorials*, vol. 18, no. 1, pp. 124–161, 2016.

[6] Q. Xu, A. Gerber, Z. M. Mao, and J. Pang, "Acculoc: Practical localization of performance measurements in 3G networks," in *Proc. ACM MobiSys*, Bethesda, Maryland, USA, 2011.

[7] I. M. Averin, V. T. Ermolayev, and A. G. Flaksman, "Locating mobile users using base stations of cellular networks," *Communications and Network*, vol. 2, no. 04, p. 216, 2010.

[8] D. Lee, S. Zhou, X. Zhong, Z. Niu, X. Zhou, and H. Zhang, "Spatial modeling of the traffic density in cellular networks," *IEEE Wireless Communications*, vol. 21, no. 1, pp. 80–88, 2014.

[9] X. Kang, L. Liu, and H. Ma, "Enhance the quality of crowdsensing for fine-grained urban environment monitoring via data correlation," *Sensors*, vol. 17, no. 1, p. 88, 2017.

[10] M. Gramaglia and M. Fiore, "Hiding mobile traffic fingerprints with glove," in *Proc. ACM CoNEXT*, Heidelberg, Germany, 2015.

[11] J. D. Simpkins and R. L. Stevenson, "An introduction to super-resolution imaging," *Mathematical Optics*, vol. 16, pp. 555–578, 2012.

[12] M. Abadi *et al.*, "TensorFlow: Large-scale machine learning on heterogeneous systems," 2015, software available from tensorflow.org. [Online]. Available: http://tensorflow.org/

[13] A. Krizhevsky, I. Sutskever, and G. E. Hinton, "Imagenet classification with deep convolutional neural networks," in *Proc. NIPS*, 2012.

[14] C. Dong, C. C. Loy, K. He, and X. Tang, "Image super-resolution using deep convolutional networks," *IEEE transactions on pattern analysis and machine intelligence*, vol. 38, no. 2, pp. 295–307, 2016.

[15] X. Yu and F. Porikli, "Ultra-resolving face images by discriminative generative networks," in *European Conference on Computer Vision*. Springer, 2016, pp. 318–333.

[16] K. He, X. Zhang, S. Ren, and J. Sun, "Deep residual learning for image recognition," in *Proceedings of the IEEE Conference on Computer Vision and Pattern Recognition*, 2016, pp. 770–778.

[17] I. Goodfellow, J. Pouget-Abadie, M. Mirza, B. Xu, D. Warde-Farley, S. Ozair, A. Courville, and Y. Bengio, "Generative adversarial nets," in *Advances in neural information processing systems*, 2014, pp. 2672–2680.

[18] H. Noh, S. Hong, and B. Han, "Learning deconvolution network for semantic segmentation," in *Proceedings of the IEEE International Conference on Computer Vision*, 2015, pp. 1520–1528.

[19] S. Ji, W. Xu, M. Yang, and K. Yu, "3D convolutional neural networks for human action recognition," *IEEE Trans. Pattern Analysis and Machine Intel.*, vol. 35, no. 1, pp. 221–231, 2013.

[20] S. Ioffe and C. Szegedy, "Batch normalization: Accelerating deep network training by reducing internal covariate shift," in *International Conference on Machine Learning*, 2015, pp. 448–456.

[21] A. L. Maas, A. Y. Hannun, and A. Y. Ng, "Rectifier nonlinearities improve neural network acoustic models," in *Proc. ICML*, vol. 30, no. 1, 2013.

[22] A. Veit, M. J. Wilber, and S. Belongie, "Residual networks behave like ensembles of relatively shallow networks," in *Advances in Neural Information Processing Systems*, 2016, pp. 550–558.

[23] G. Huang, Z. Liu, K. Q. Weinberger, and L. van der Maaten, "Densely connected convolutional networks," *To appear Proceedings of the IEEE Conference on Computer Vision and Pattern Recognition*, 2017.

[24] K. Simonyan and A. Zisserman, "Very deep convolutional networks for large-scale image recognition," *International Conference on Learning Representations (ICLR)*, 2015.

[25] C. Ledig, L. Theis, F. Huszár, J. Caballero, A. Cunningham, A. Acosta, A. Aitken, A. Tejani, J. Totz, Z. Wang *et al.*, "Photo-realistic single image super-resolution using a generative adversarial network," *To appear Proceedings of the IEEE Conference on Computer Vision and Pattern Recognition*, 2017.

[26] I. Goodfellow, Y. Bengio, and A. Courville, *Deep Learning*. MIT Press, 2016, http://www.deeplearningbook.org.

[27] M. Arjovsky and L. Bottou, "Towards principled methods for training generative adversarial networks," in *NIPS 2016 Workshop on Adversarial Training*, vol. 2016.

[28] D. Kingma and J. Ba, "Adam: A method for stochastic optimization," *International Conference on Learning Representations (ICLR)*, 2015.

[29] G. Barlacchi, M. De Nadai, R. Larcher, A. Casella, C. Chitic, G. Torrisi, F. Antonelli, A. Vespignani, A. Pentland, and B. Lepri, "A multi-source dataset of urban life in the city of Milan and the province of Trentino," *Scientific data*, vol. 2, 2015.

[30] R. Carlson and F. Fritsch, "Monotone piecewise bicubic interpolation," *SIAM journal on numerical analysis*, vol. 22, no. 2, pp. 386–400, 1985.

[31] J. Yang, J. Wright, T. S. Huang, and Y. Ma, "Image super-resolution via sparse representation," *IEEE transactions on image processing*, vol. 19, no. 11, pp. 2861–2873, 2010.

[32] R. Timofte, V. De Smet, and L. Van Gool, "A+: Adjusted anchored neighborhood regression for fast super-resolution," in *Asian Conference on Computer Vision*. Springer, 2014, pp. 111–126.

[33] "TensorLayer library," accessed Jan 2017. [Online]. Available: https://tensorlayer.readthedocs.io/en/latest/

[34] T. Chai and R. R. Draxler, "Root mean square error (RMSE) or mean absolute error (MAE)?–Arguments against avoiding RMSE in the literature," *Geoscientific Model Development*, vol. 7, no. 3, pp. 1247–1250, 2014.

[35] A. Hore and D. Ziou, "Image quality metrics: PSNR vs. SSIM," in *Proc. IEEE International Conference on Pattern Recognition (ICPR)*, 2010, pp. 2366–2369.

[36] J. Li, X. Chen, E. Hovy, and D. Jurafsky, "Visualizing and understanding neural models in nlp," in *Proceedings of NAACL-HLT*, 2016, pp. 681–691.

[37] M. Roughan, Y. Zhang, W. Willinger, and L. Qiu, "Spatio-temporal compressive sensing and internet traffic matrices," *IEEE/ACM Transactions on Networking (ToN)*, vol. 20, no. 3, pp. 662–676, 2012.

[38] L. Kong, M. Xia, X.-Y. Liu, G. Chen, Y. Gu, M.-Y. Wu, and X. Liu, "Data loss and reconstruction in wireless sensor networks," *IEEE Transactions on Parallel and Distributed Systems*, vol. 25, no. 11, pp. 2818–2828, 2014.

[39] L. Liu, W. Wei, D. Zhao, and H. Ma, "Urban resolution: New metric for measuring the quality of urban sensing," *IEEE Transactions on Mobile Computing*, vol. 14, no. 12, pp. 2560–2575, 2015.

[40] C. E. Duchon, "Lanczos filtering in one and two dimensions," *Journal of Applied Meteorology*, vol. 18, no. 8, pp. 1016–1022, 1979.

[41] R. Timofte, V. De Smet, and L. Van Gool, "Anchored neighborhood regression for fast example-based super-resolution," in *Proc. IEEE International Conference on Computer Vision*, 2013, pp. 1920–1927.